\documentclass{aastex}          
\usepackage{spr-astr-addons}    

\begin{document}
%
\title{First frequency analysis for three new members of the group of eclipsing binaries with a pulsating component}

\shorttitle{Frequency analysis for three EBs with a pulsating component}

\shortauthors{Liakos \& Caga\v{s}}

\author{A. Liakos\altaffilmark{1}}
\email{alliakos@noa.gr} 
\and
\author{P. Caga\v{s}\altaffilmark{2}}

\altaffiltext{1}{National Observatory of Athens, Institute for Astronomy, Astrophysics, Space Applications and Remote Sensing, I. Metaxa \& Pavlou St., GR-152 36, Penteli, Athens, Hellas}
\altaffiltext{2}{BSObservatory, Modr\'{a} 587, 760 01 Zl\'{\i}n, Czech Republic}

\begin{abstract}
We present the first light curves and pulsation analysis results for V729~Aql and two newly discovered eclipsing binaries, namely USNO-A2.0~0975-17281677 and USNO-A2.0~1200-03937339. Frequency search was applied on the residuals of their light curves and the results showed that their primary components pulsate in multiperiodic modes and lie well inside the frequency and temperature range of $\delta$~Scuti stars. Moreover, for USNO-A2.0~1200-03937339 two frequencies inside the $\gamma$~Dor frequency range were also detected, but their origin is discussed. The photometric models of USNO-A2.0~1200-03937339 and V729~Aql are also presented, while their absolute parameters as well as the evolutionary status of their components were roughly estimated.

\end{abstract}

\keywords{Methods: data analysis -- Methods: observational -- stars: binaries:eclipsing -- stars: fundamental parameters -- stars: variables: $\delta$ Scuti -- Stars: evolution -- stars:individual: V729~Aql, USNO-A2.0~0975-17281677, USNO-A2.0~1200-03937339}

\section{Introduction}
\label{INTRO}

In general, on one hand, eclipsing binary systems (hereafter EBs) provide the means for the calculation of stellar absolute parameters and evolutionary status. On the other hand, pulsating stars reveal the `secrets' of the stellar interior with their oscillating behaviour. Thus, the cases of binaries with a $\delta$~Scuti component can be considered as extremely important because they combine information of two different types of stellar variability, which is very useful for the complete study of interacting stellar evolution.

\citet{MK02} firstly suggested that the binary systems with mass gain pulsators must be considered as a new category of variables, which they named it as `oscillating eclipsing Algol-type stars' (oEA stars). Up to date, only $\sim 85$ binaries containing a $\delta$~Scuti component have been discovered, but their number is rapidly increasing, especially due to the large amount of data from space telescopes (i.e. \textit{Kepler}, \textit{Corot}). \citet{SO06b} published the first list of candidate systems of this type, while \citet{SO06a} made a first attempt to find a correlation between orbital ($P_{\rm orb}$) and pulsation ($P_{\rm puls}$) periods for these systems. \citet{LI12}, after a 5~yr survey on these candidates \citep[cf.][]{LN09,LN13}, based on a larger sample, found a more accurate connection between $P_{\rm orb}$-$P_{\rm puls}$ and also other correlations between the properties of pulsators and their host systems. Up to 2013, only works based on observational data had been published, but \citet{ZH13} presented the first theoretical establishment for the $P_{\rm orb}-P_{\rm puls}$ connection.


\begin{table*}
\centering
\caption{Detailed observations log (N=number of nights, DP=data points collected, sd=standard deviation in mmag).}
\label{tabobslog}
\scalebox{0.9}{
\begin{tabular}{l c l cc ccc}
\tableline																				
System	                &	   N	& Observation dates (YYYY:DD/MM)	          &     DP     &$sd$ &       Comparison 	 &$m_{\rm V}^{\rm b}$ &       $B-V^{\rm b}$     \\
\tableline																					
0975-17281677$^{\rm a}$	&	   13	&2011: 21/10; 2, 10/11                        &      763   & 4.0 &0975-17288906$^{\rm a}$&          12.8      & 0.26              \\
                        &           &2012: 3, 11, 19, 20, 22/10; 14, 15/11        &                                     &&&&
                  \\
                        &           &2013: 3, 5, 7/9                              &                                     &&&&
                        \\
\tableline
1200-03937339$^{\rm a}$	&	   26	&2012: 30, 31/1; 1, 10, 12, 26/2; 3, 5, 6/3   &     1775   & 8.2 &1200-03910760$^{\rm a}$&          12.4      &  0.44                    \\
                        &           &2012: 20, 21/10; 14, 15/11                   &                                     &&&&                                      \\
                        &           &2013: 27, 28/2; 2, 4/3; 1/4; 27/12           &                                     &&&&                                      \\
                        &           &2014: 3, 12, 26/1; 3, 4, 6, 20/2             &                                     &&&&                                      \\
\tableline																					
V729~Aql            	&	   19	&2011: 29/9; 22/10; 2, 3, 9, 10/11            &     1049   & 6.7 &0975-17188972$^{\rm a}$ 	&       14.1     & 0.16    \\
                        &           &2012: 3, 9, 11, 19, 20, 21, 22/10; 14, 15/11 &                                     &&&&                                      \\
                        &           &2013: 16/8; 3, 5, 7/9                        &                                     &&&&                                      \\
\tableline																				
\multicolumn{8}{l}{$^{\rm a}USNO-A2.0$ catalogue \citep{MO98}; $^{\rm b}NOMAD$ catalogue \citep{ZA04}}									
\end{tabular}}
\end{table*}

In the present work, we firstly report the discoveries of USNO-A2.0 0975-17281677 (= UCAC4~515-117553 = 2MASS~19573100+1254164 = CzeV331~Aql, $\alpha_{2000}=19^{\rm h}~57^{\rm m}~30.96^{\rm s}$ and $\delta_{2000}=+12^{\circ}~54'~17.3''$) and USNO-A2.0~1200-03937339 (= UCAC4 605-026193 = 2MASS~05515747+3054538 = CzeV354 Aur, $\alpha_{2000}=05^{\rm h}~51^{\rm m}~57.4^{\rm s}$ and $\delta_{2000}=+30^{\circ}~54'~54.1''$) as eclipsing systems and then we focus on their pulsating components. The only available information for these objects comes from general catalogues. In particular, for USNO-A2.0 0975-17281677 the \textit{UCAC4} catalogue \citep{ZA13} gives the magnitude in $V$-filter as 13.497 and a $B-V$ value of 0.452, while the \textit{2MASS All-Sky Catalog of Point Sources} \citep{SK06} gives $J-H=0.382$. For USNO-A2.0 1200-03937339 the same catalogues refer $m_{\rm V}$=14.528, $B-V$=0.354, and $J-H=0.2$, respectively.

V729~Aql \citep[$m_{\rm V}$=13.756--\textit{UCAC4} catalogue;][]{ZA13}  was discovered as a variable with a period of $\sim1.28^{\rm d}$ by \citet{HO44}, but no other studies exist so far. The only information for this system comes from \citet{SK90}, who calculated its mass ratio as 0.52 and classified its components as A3 and G0IV. The \textit{2MASS All-Sky Catalog of Point Sources} \citep{SK06} gives $J-H=0.198$, while the $Lick~NPM2$ \citep{HA04} and the $UCAC4$ \citep{ZA13} catalogues give $B-V$ values of 0.36 and 0.39, respectively.

For V729~Aql and USNO-A2.0~1200-03937339 a light curve (hereafter LC) modeling, as well as, a rough estimation for their absolute parameters and their present evolutionary status are presented. For all systems, we used Fourier transformation methods on their LC residuals after subtracting the binarity effects to detect the oscillating frequencies of their pulsating components. Finally, these pulsators are compared with others with similar properties. In all following tables formal errors are indicated in parentheses along-side adopted values.

\section{Observations, data reduction, and ephemerides' calculation}
\label{OBS}

The observations of the systems (available online at: http://var2.astro.cz/EN/obslog.php under `Pavel Caga\v{s}') were carried out at the BSObservatory (Zl\'{\i}n, Czech Republic) with a 25~cm Netwonian reflector (f/5.4) equipped with the G4-16000 CCD camera. The field-of-view (FoV) for this setup was $71'\times71'$. As the aim of these measurements was searching for new variables, observations were made without filtering to gain the highest possible throughput, therefore the following results concern `white'-observations. Individual exposures were 240~s or 180~s long depending on seeing and transparency of the particular observing night. Differential aperture photometry as well as the calibrations were performed using the \textsc{C-munipack} software package \citep{HR98,MO04}. For each system the selected comparison star was one of the several carefully chosen potential ones in the FoV. Its selection was based on its $B-V$ index, in order to be close to the index of individual observed star and thus to correct for atmospheric extinction. Given that the FoVs of all targets were quite dense, we carefully examined the possibility of blend on each one. For USNO-A2.0 1200-03937339 we examined the star shape in both our frames ($1.39''$/px) as well as in the publicly available SDSS DR7 data \citep{AB09} with $0.396''$/px in $u$ and $z$ filters, but we did not find any star image deformations hinting a possible blend. On the other side, the shape of USNO-A2.0 0975-17281677 is obviously prolonged on images from our telescope, indicating another star is at least partially projected to the photometric aperture. However, we repeated the photometry with various small enough apertures (1-3 px) to make sure that both Algol-type and $\delta$~Scuti-type variations originate in the single star only. The shape of V729~Aql is rounded without any visible trace of another close star on our images. In Table~\ref{tabobslog} we present the detailed observations log, which contains: the total nights of observations and their dates, the data points acquired and their mean photometric error (standard deviation), and the names as well as a few properties of the comparison stars used.

We used the software \textsc{minima}~v2.3 (Nelson 2009) to calculate the times of minima which are listed in Table~\ref{tabtom}. For USNO-A2.0 0975-17281677 the minimum timings are not enough to determine its period with high accuracy, however, their time difference sets an upper limit of 12.0664~days. Using this value and all its integer submultiples for phasing the data we found that the most possible period value is 3.0155~days. Finally, we derived the first preliminary ephemeris of the system as:
\begin{equation}
Min.I~(HJD)= 2455868.3237(5)+3.0155(8)^{d}\times E
\end{equation}
For USNO-A2.0 1200-03937339 the amount of minimum timings was sufficient for deriving its ephemeris. Thus, using the least squares method on these minima we found the following ephemeris:
\begin{equation}
Min.I~(HJD)= 2455958.4365(2)+1.17962(1)^{d}\times E
\end{equation}
Finally, for V729~Aql we used our calculated time of minimum as reference and the literature period for the derivation of its ephemeris:
\begin{equation}
Min.I~(HJD)= 2455876.3417(4)+1.281905^{d}\times E
\end{equation}

\begin{table}[t]
\centering
\caption{Times of minima based on our observations.}
\label{tabtom}
\scalebox{0.85}{
\begin{tabular}{cc cc}																					
\tableline
 HJD$-$2,400,000.0  &       Type             &      HJD$-$2,400,000.0   &       Type      \\
\tableline
\multicolumn{2}{c}{USNO-A2.0~0975-17281677}  &\multicolumn{2}{c}{USNO-A2.0~1200-03937339} \\
\tableline
55856.2573 (7)      &     I                  &      55958.4367 (4)	    &	I	          \\
55868.3237 (5)      &     I                  &      55984.3878 (4)      &	I	          \\
\cline{1-2}
\multicolumn{2}{c}{V729~Aql}                 &      55990.2864 (3)	    &	I	          \\
\cline{1-2}
55876.3417 (4)      &     I 				 &      56221.4932 (3)	    &	I	          \\
                    &                        &      56693.3427 (2)	    &	I	          \\
\tableline						
\end{tabular}}
\end{table}

\section{Light curve modeling}
\label{LCM}

\begin{figure}[h!]
\centering
\begin{tabular}{c}
\includegraphics[width=7cm]{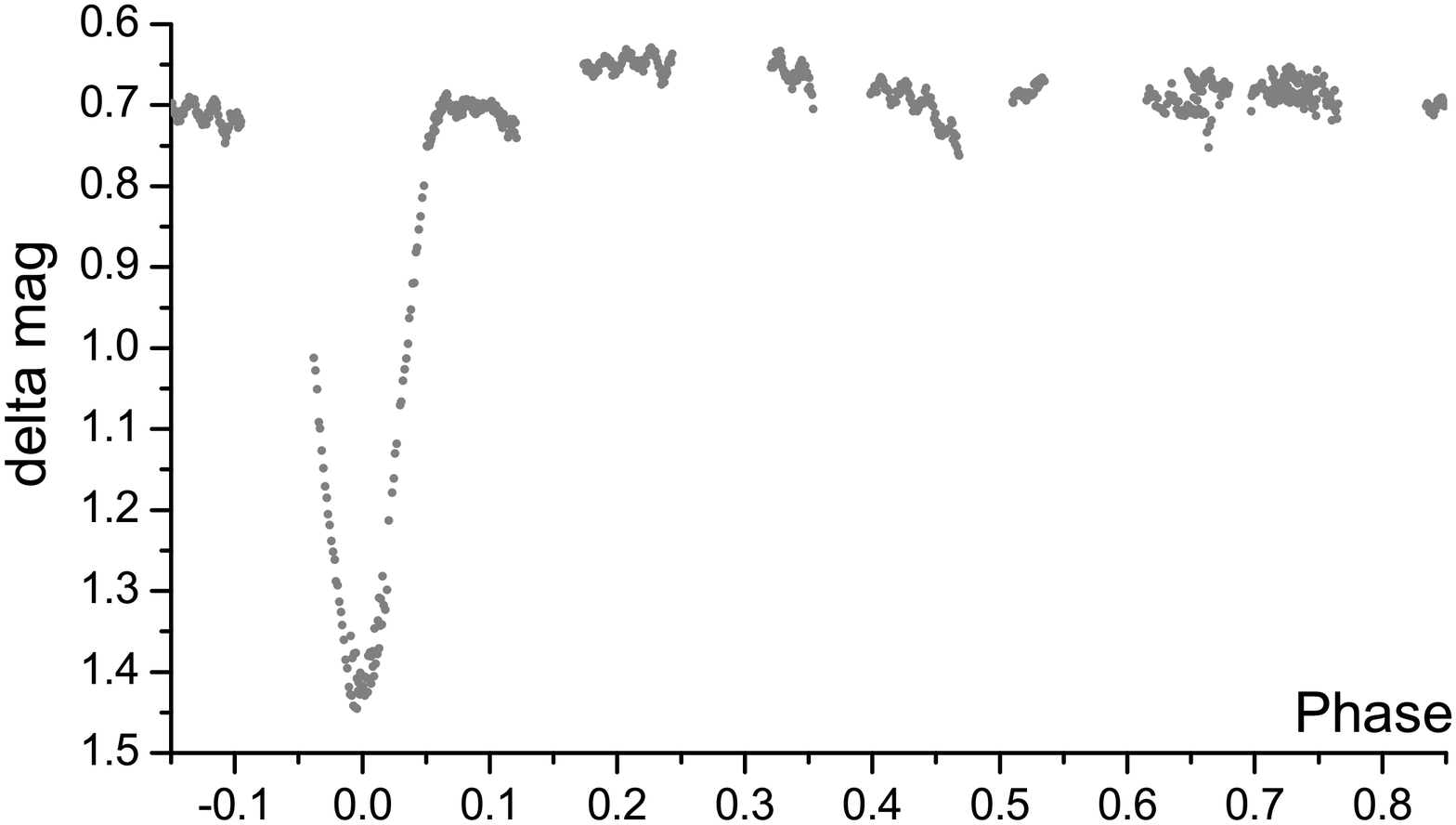}\\
\includegraphics[width=7cm]{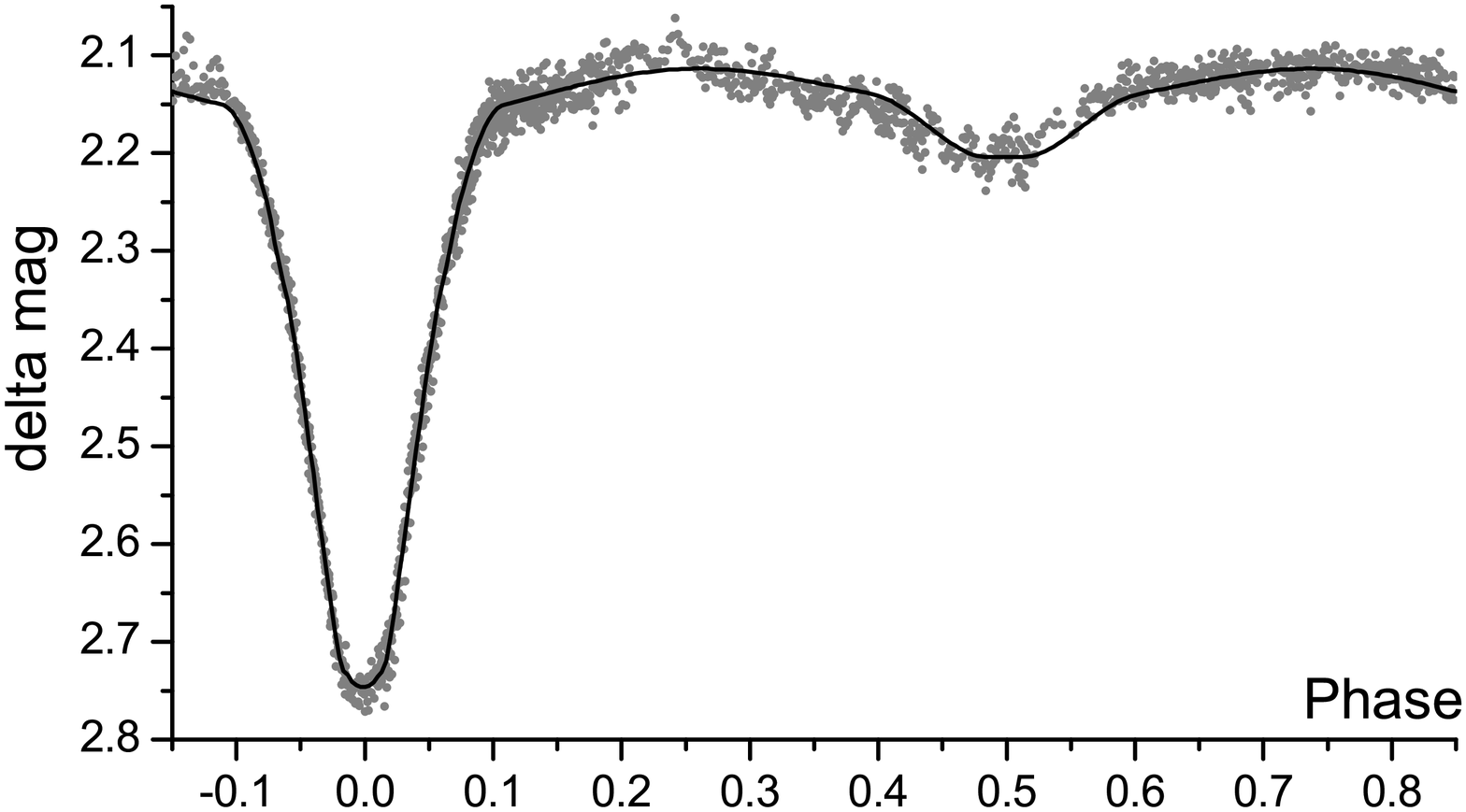}\\
\includegraphics[width=7cm]{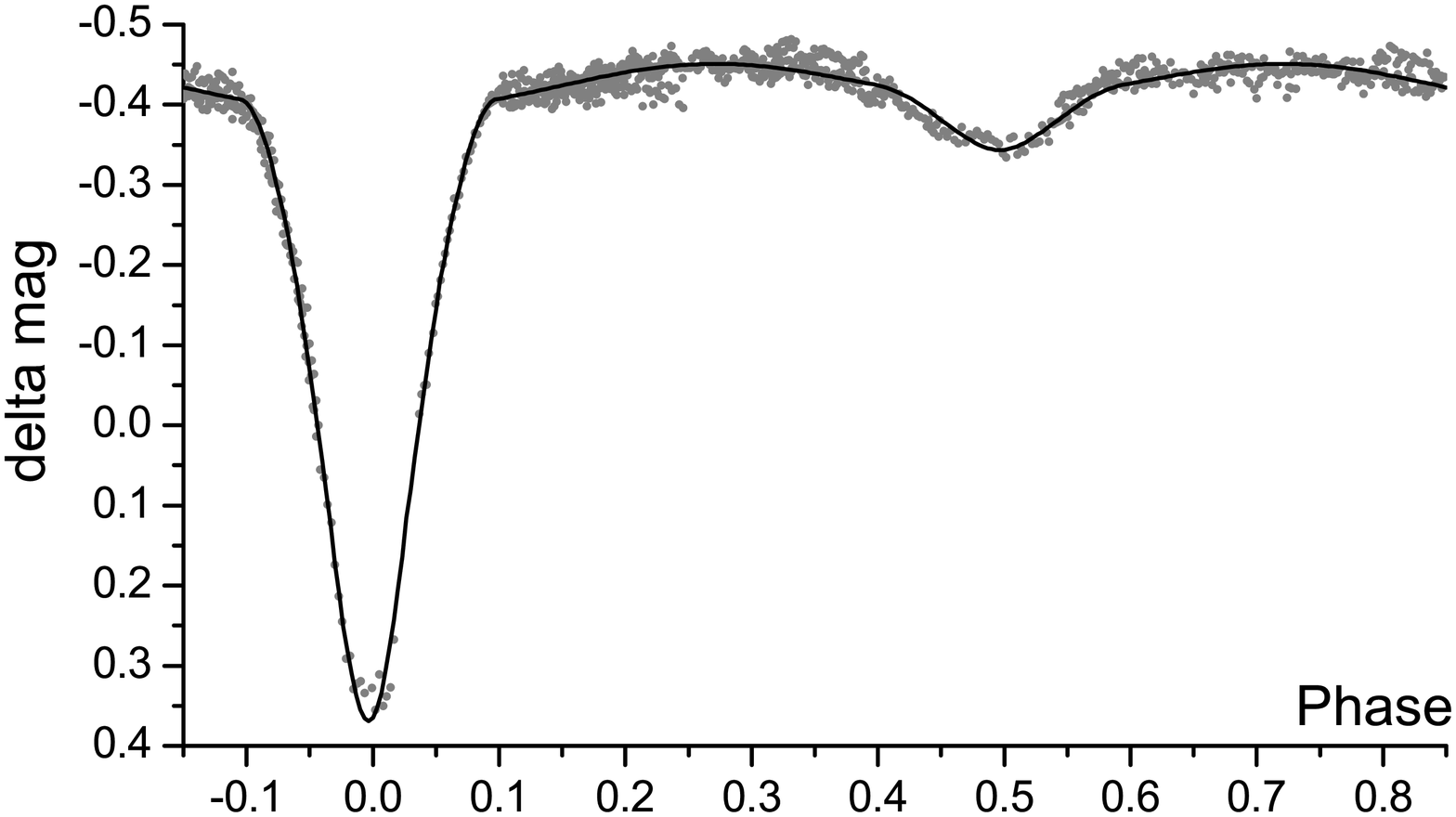}\\
\end{tabular}
\caption{Observed LC of USNO-A2.0 0975-17281677 (top), and observed (points) and synthetic (line) LCs of USNO-A2.0~1200-03937339 (middle), and V729~Aql (bottom).}
\label{figLCs}
\end{figure}

Unfortunately, $\sim38\%$ of the LC of USNO-A2.0 0975-17281677 was not covered by our observations. In particular, the critical phaseparts of the deepest part of the secondary minimum and the beginning of the descending part of the primary eclipse are totally missing. For this reason, we decided not to present any photometric model. However, the phased data based on the preliminary ephemeris (see section~\ref{OBS}) are plotted in Fig.~\ref{figLCs}.

On the other hand, full LCs for USNO-A2.0~1200-03937339 and V729~Aql were obtained. The LC of each system was analysed using all individual observations with the \emph{PHOEBE} v.0.29d software \citep{PZ05}, that is based on the Wilson-Devinney (W-D) code \citep{WD71,WI79,WI90}. In the absence of spectroscopic mass ratio, the `$q$-search' method with a step of 0.1 \citep[cf.][]{LN12} was applied in modes 2 (detached system), 4 (semi-detached binary with its primary component filling its Roche lobe) and 5 (conventional semi-detached binary) to find feasible (`photometric') estimates for the mass ratio. After the finding of the most possible $q$, this value was set as adjustable parameter in the subsequent analysis for further converging. The (linear) limb darkening coefficients, $x_1$ and $x_2$, were taken from \citet{VH93}; the dimensionless potentials $\Omega_{1}$ and $\Omega_{2}$, the fractional luminosity of the primary component $L_{1}$, and the system's orbital inclination $i$ were set as adjustable parameters. Given that our data are `white', the filter depended parameters (i.e. $L$ and $x$) correspond to bolometric values ($bol$).

The temperatures of the primary components ($T_{1}$) were assumed according to the colour indices of the systems as given in the literature (see section~\ref{INTRO}) based on the colour index-temperature-spectral type correlations of \citet{CO00}. Therefore, initially a primary's temperature $T_{1}$ of 7000~K (F2 spectral class) for USNO-A2.0 1200-03937339 and 6900~K (F3 spectral class) for V729~Aql were set, while the temperatures of secondaries ($T_2$) were adjusted. However, the distance of USNO-A2.0~1200-03937339 has been estimated as 2300~pc \citep{PD10}, therefore its colour indices might have been affected by interstellar reddening. For this reason, we made various models for the system inside the temperature range $7000\pm500$~K with a step of 250~K. For each model we took into account the change of the bolometric albedos $A$ and gravity darkening coefficients $g$. Particulary, for radiative stellar atmospheres we assumed $A$=1 and $g$=1 \citep{RU69,VZ24}, while for convective ones $A$=0.5 and $g$=0.32 \citep{RU69,LU67}. Although a distance measurement is not available for V729~Aql, we followed the same procedure in the range $6900\pm500$~K with the same step.

The conventional semi-detached Roche geometry was found as the most appropriate mode for both systems. The final solution of each one corresponds to the model having the minimum sum of the squared residuals of the individual models made regarding the assumed temperature and the atmosphere type of the primary component. In Fig.~\ref{figLCs} the observed points as well as the theoretical lines are plotted, while in Table~\ref{tabLCM} are given the models' parameters.

\begin{table}[h!]
\centering
\caption{Light curve parameters for USNO-A2.0~1200-03937339 and V729~Aql. }
\label{tabLCM}
\scalebox{0.78}{
\begin{tabular}{l cc cc}
\tableline																					
Parameter	                   &                      \multicolumn{4}{c}{Value}	                  \\
\tableline	
                &  \multicolumn{2}{c}{USNO-A2.0~1200-03937339}    & \multicolumn{2}{c}{V729~Aql}  \\		
\tableline									
$i~(^\circ$)	               &  \multicolumn{2}{c}{84.6 (2)}    & \multicolumn{2}{c}{77.3 (2)}  \\
$q~(m_{2}/m_{1}$)	           &  \multicolumn{2}{c}{0.19 (2)}    & \multicolumn{2}{c}{0.44 (1)}  \\
\tableline																					
$Component$	                   &        Primary   & Secondary     &        Primary   & Secondary  \\
\tableline																					
$T$ (K)                        &	7250$^{\rm a}$&    4320 (108) &	6900$^{\rm a}$&    4300 (175) \\
$A^{\rm a}$                    &        1         &     0.5       &      0.5      &       0.5     \\
$g^{\rm a}$                    &        1         &     0.32      &      0.32     &       0.32    \\
$\Omega$	                   &	2.91 (1)	  &	2.22$^{\rm b}$&	3.84 (3)	  &	2.77$^{\rm b}$\\
$x_{\mathrm{bol}}$	           &	    0.483	  &	   0.823	  &	    0.500	  &	   0.822      \\
$(L/L_\mathrm{T})_\mathrm{bol}$&	   0.964 (2)  &	   0.036 (1)  &	   0.903 (4)  &	   0.097 (1)  \\
$r_\mathrm{pole}$              &       0.366 (1)  &    0.230 (1)  &    0.293 (1)  &    0.291 (1)  \\
$r_\mathrm{point}$             &       0.386 (1)  &    0.337 (1)  &    0.306 (1)  &    0.418 (1)  \\
$r_\mathrm{side}$              &       0.377 (1)  &    0.239 (1)  &    0.299 (1)  &    0.303 (1)  \\
$r_\mathrm{back}$              &       0.382 (1)  &    0.271 (1)  &    0.303 (1)  &    0.336 (1)  \\
\tableline																					
\multicolumn{5}{l}{$^{\rm a}$assumed, $^{\rm b}$fixed, $L_{\mathrm{T}}= L_1+L_2$, $r$=relative radius}\\																				
\end{tabular}}
\end{table}

\section{Absolute parameters}
\label{ABS}
\begin{figure}[h!]
\centering
\begin{tabular}{cc}
            \multicolumn{2}{c}{\includegraphics[width=6.7cm]{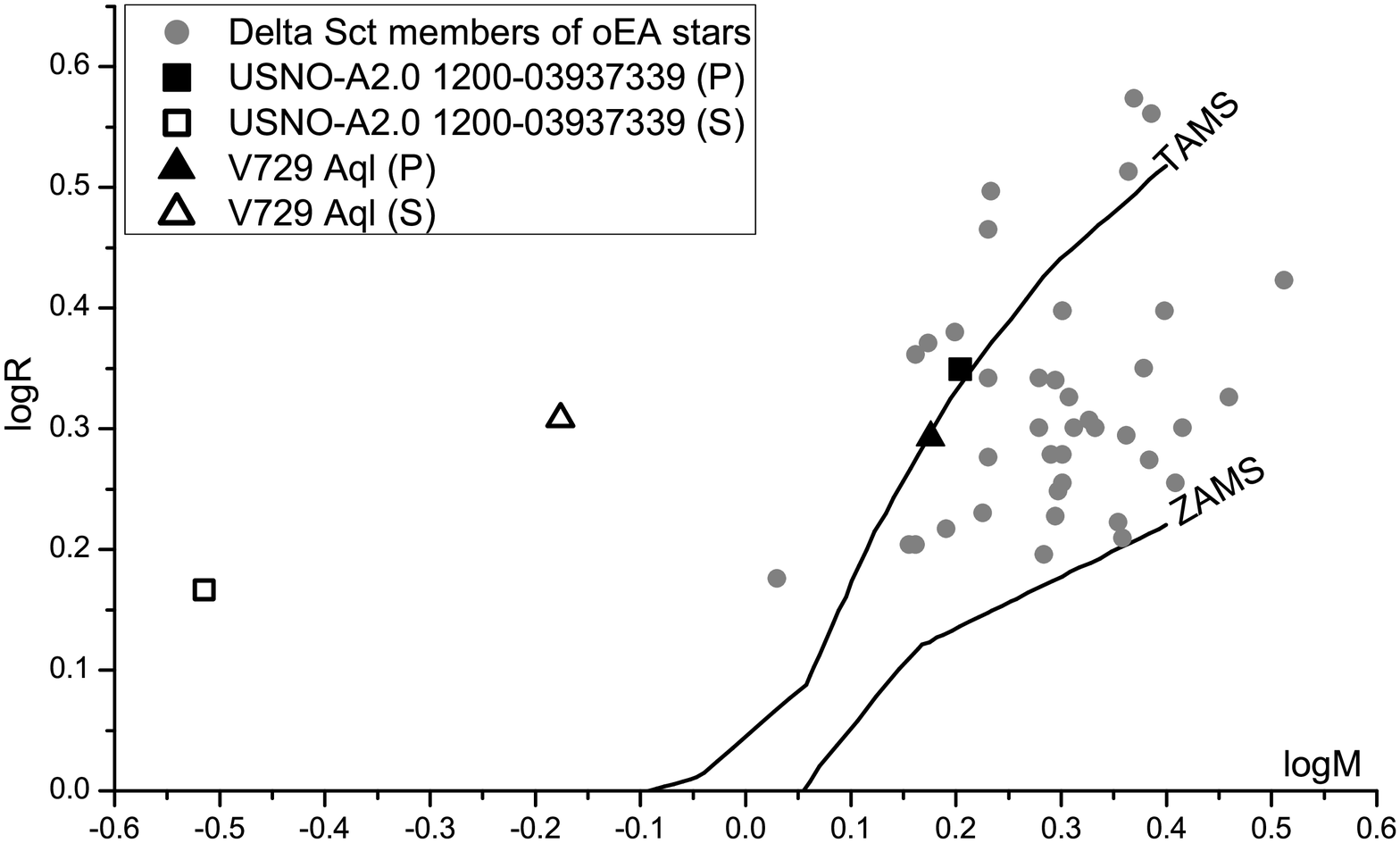}}                \\
\includegraphics[width=3.2cm]{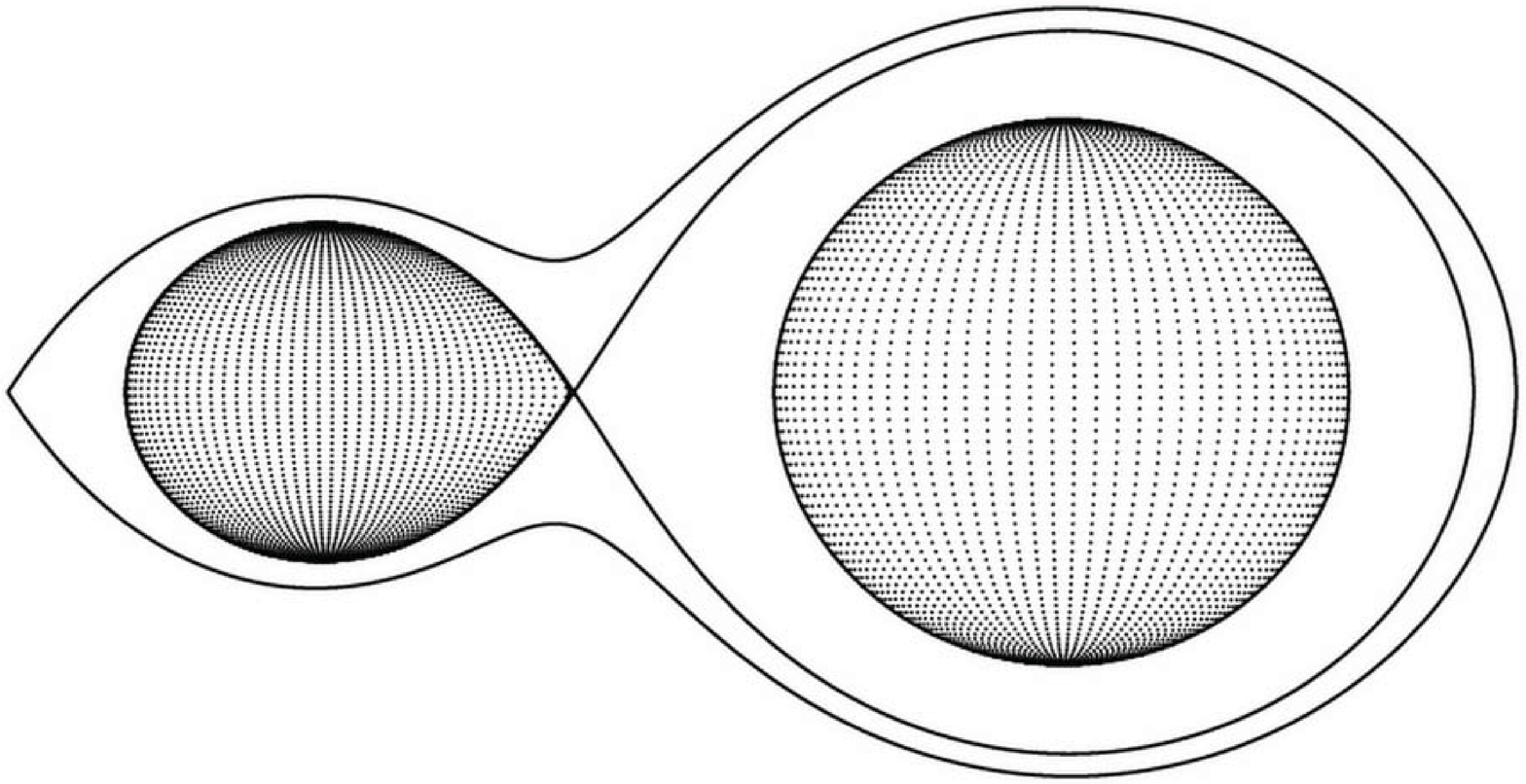}&\includegraphics[width=3.2cm]{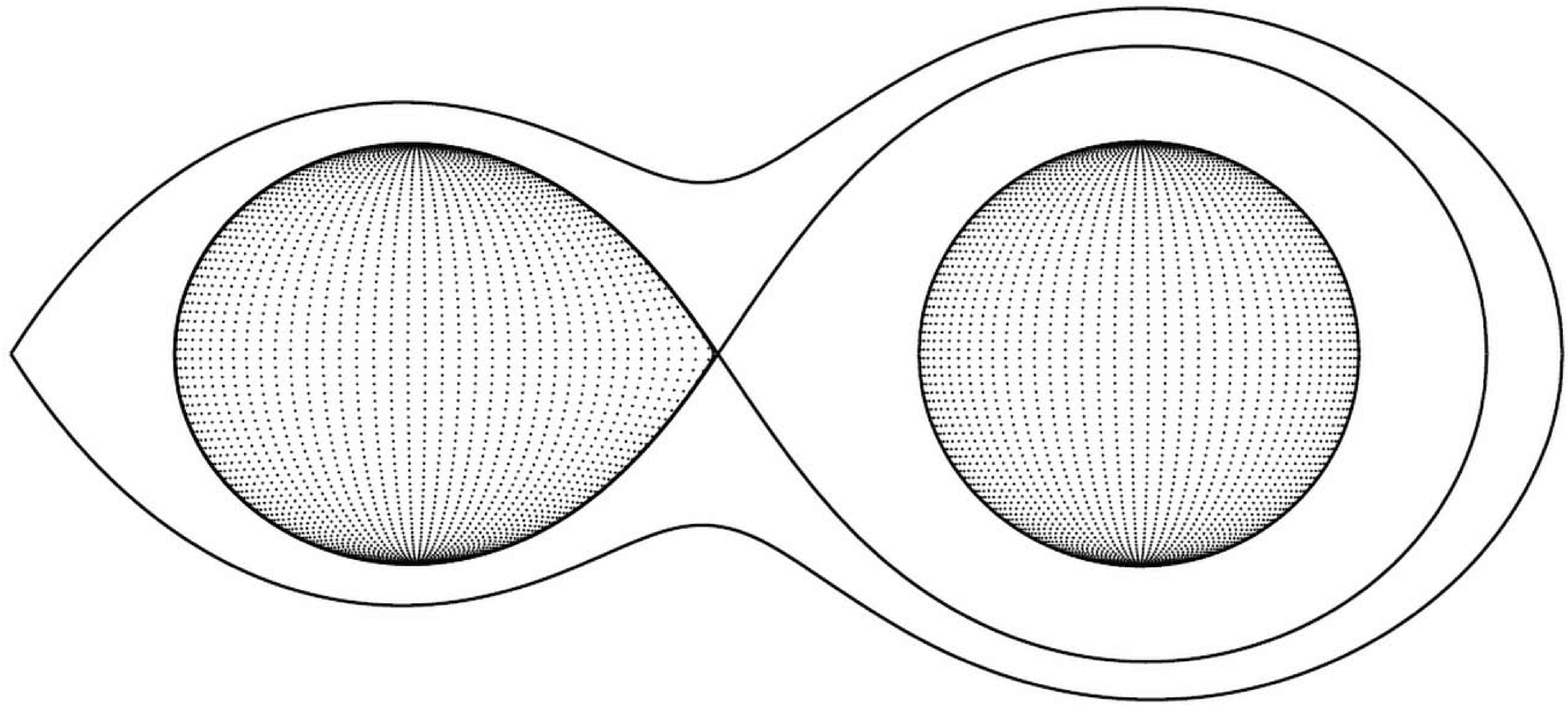}\\
\end{tabular}
\caption{Top: The positions of the components (P=primary, S=secondary) of USNO-A2.0 1200-03937339 and V729~Aql in the $M-R$ diagram along with the $\delta$~Scuti stars-members of oEA stars. Bottom: The three-dimensional views (left star=secondary component) of USNO-A2.0 1200-03937339 (left) and V729~Aql (right).}
\label{figMR3D}
\end{figure}

Although no radial velocity measurements exist for USNO-A2.0 1200-03937339 and V729~Aql, we can form fair estimates of the absolute parameters of their components. The masses of the primaries were assumed according to their spectral types using the correlations of \citet{CO00}, while a fair error of 10\% of the mass value was also assumed in order to obtain more realistic conclusions. The secondaries' masses follow from the determined mass ratios (see section~\ref{LCM}) and the semi-major axes $a$ are then derived from Kepler's third law. The errors were calculated using the error propagation method. The parameters are listed in Table~\ref{tabAbs}, while the positions of the systems' components in the Mass-Radius ($M-R$) diagram as well as a 3-D representation of their Roche geometry are given in Fig.~\ref{figMR3D}. The theoretical lines for Zero Age Main Sequence (ZAMS) and Terminal Age Main Sequence (TAMS) were taken from \citet{NM03}, while the data for the $\delta$~Scuti stars-members of oEA stars were taken from \citet{LI12}.

Both systems according to the absolute parameters of their components and their Roche geometry can be considered as classical Algols. Their secondary components, although they are less massive, they were found to be more evolved than the primaries, which is typical for interacting stars with past mass exchange. In Fig.~\ref{figMR3D} it can be noticed that the primaries of both systems are located very close to the TAMS limit. Both stars follow well the distribution of the so far known $\delta$~Sct stars of oEA systems. Additionally, judging from the spectral type and the rest properties of these stars (i.e. logg, position in $M-R$ diagram), we can plausibly conclude that these are the pulsating components of the systems and, moreover, that the systems may be classified as oEA stars (see section~\ref{INTRO}).

\begin{table}[h]
\centering
\caption{Absolute parameters of the components of USNO-A2.0~1200-03937339 and V729~Aql.}
\label{tabAbs}
\scalebox{0.75}{
\begin{tabular}{l cc cc}
\tableline		
Parameter         &                         \multicolumn{4}{c}{Value}	                             \\
\tableline	
		          &  \multicolumn{2}{c}{USNO-A2.0~1200-03937339}    & \multicolumn{2}{c}{V729~Aql}   \\																
\tableline	
$Component$	      &        Primary   & Secondary &        Primary   & Secondary                      \\
\tableline		
$M$~(M$_{\sun}$)  &1.6 (2)$^{\rm a}$ &	0.3 (1)  &1.5 (2)$^{\rm a}$ &	0.7 (1)		                 \\
$R$~(R$_{\sun}$)  &	  2.24 (4)	    &  1.44 (3)	 &   1.96 (1)       &	2.03 (1) 	                 \\
$L$~(L$_{\sun}$)  &	   12.4 (5)	    &	0.7 (1)	 &    7.8 (1)	    &	1.3 (2) 	                 \\
$\log g$~(cm/s$^2$)&   3.9 (1)	    &	3.6 (1)	 &	  4.0 (1)	    &	3.7 (1)	                     \\
$a$~(R$_{\sun}$)  &	   1.0 (1)	    &	5.0 (1)  &	  2.0 (1)	    &	4.6 (1)	                     \\
\tableline																					
\multicolumn{5}{l}{$^{\rm a}$assumed}
\end{tabular}}
\end{table}

\section{Frequency analysis}
\label{PULS}

For USNO-A2.0 0975-17281677, although the photometric model is substantial for pulsation analysis, we applied the following method to subtract the effects of binarity and for isolating the pulsations: We fitted polynomials (of 1st or 2nd order) on the data of each individual night, creating theoretical points, which then were subtracted from the respective observed ones. Therefore, for each night we derived the residuals after the subtraction of any other variation except the pulsations. This method is sufficient only for the detection of pulsations with periods less than the observation duration (i.e. a few hours). On the other hand, for USNO-A2.0 1200-03937339 and V729~Aql the theoretical points (LC model) were subtracted from the observed ones in order to create the binary-free LC residuals.

\begin{figure}[h!]
\centering
\begin{tabular}{c}
\includegraphics[width=5.3cm]{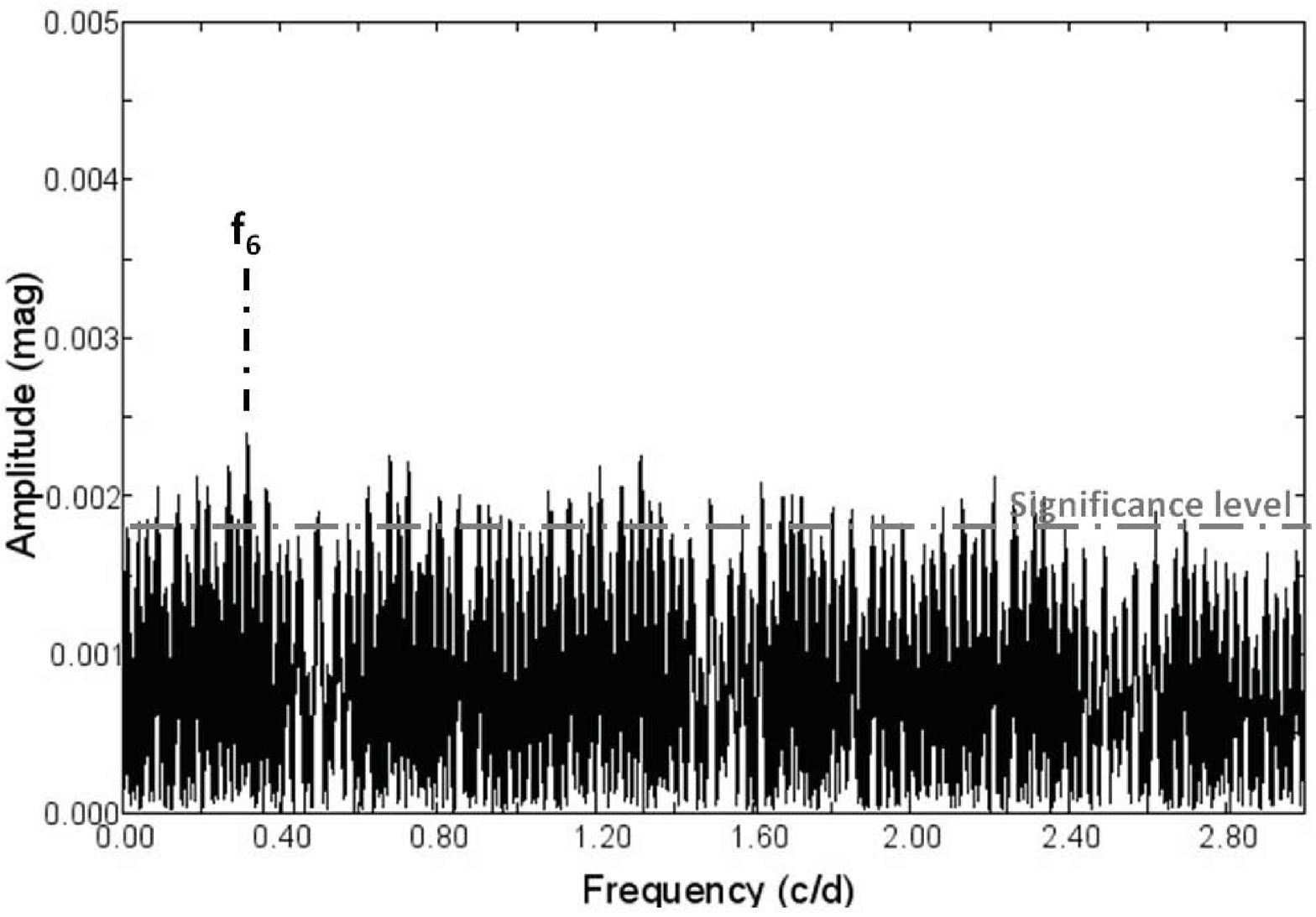}\\
\includegraphics[width=5.3cm]{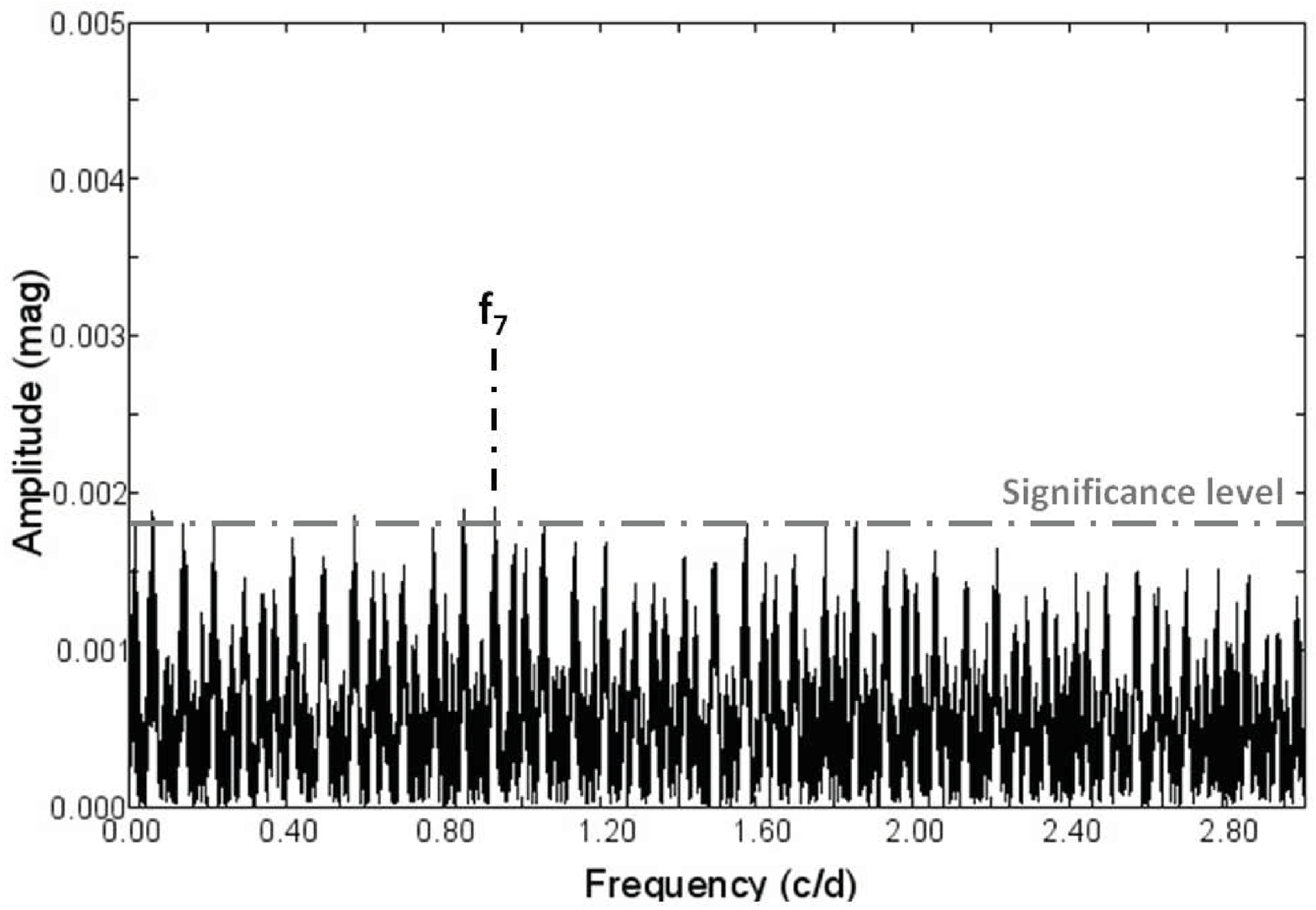}\\

\end{tabular}
\caption{Periodograms of the detected frequencies inside the $\gamma$~Dor frequency range for USNO-A2.0 1200-03937339. The significance level ($\sim0.0017$~mag) is indicated.}
\label{figfreqs_gdor}
\end{figure}

\begin{figure*}
\centering
\begin{tabular}{ccc}
        USNO-A2.0~0975-17281677           &          USNO-A2.0~1200-03937339         &               V729~Aql                   \\
\includegraphics[width=5.3cm]{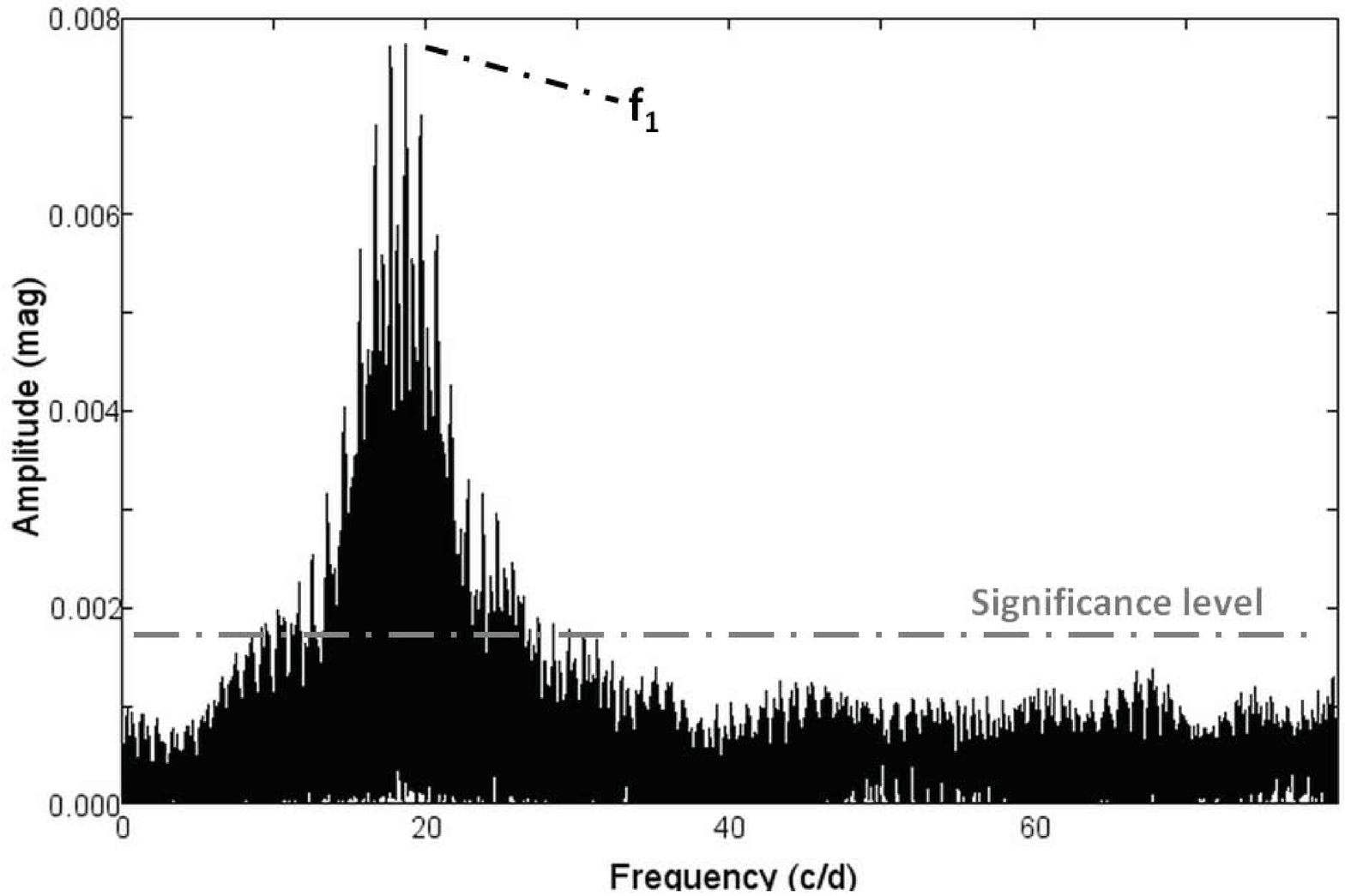}&\includegraphics[width=5.3cm]{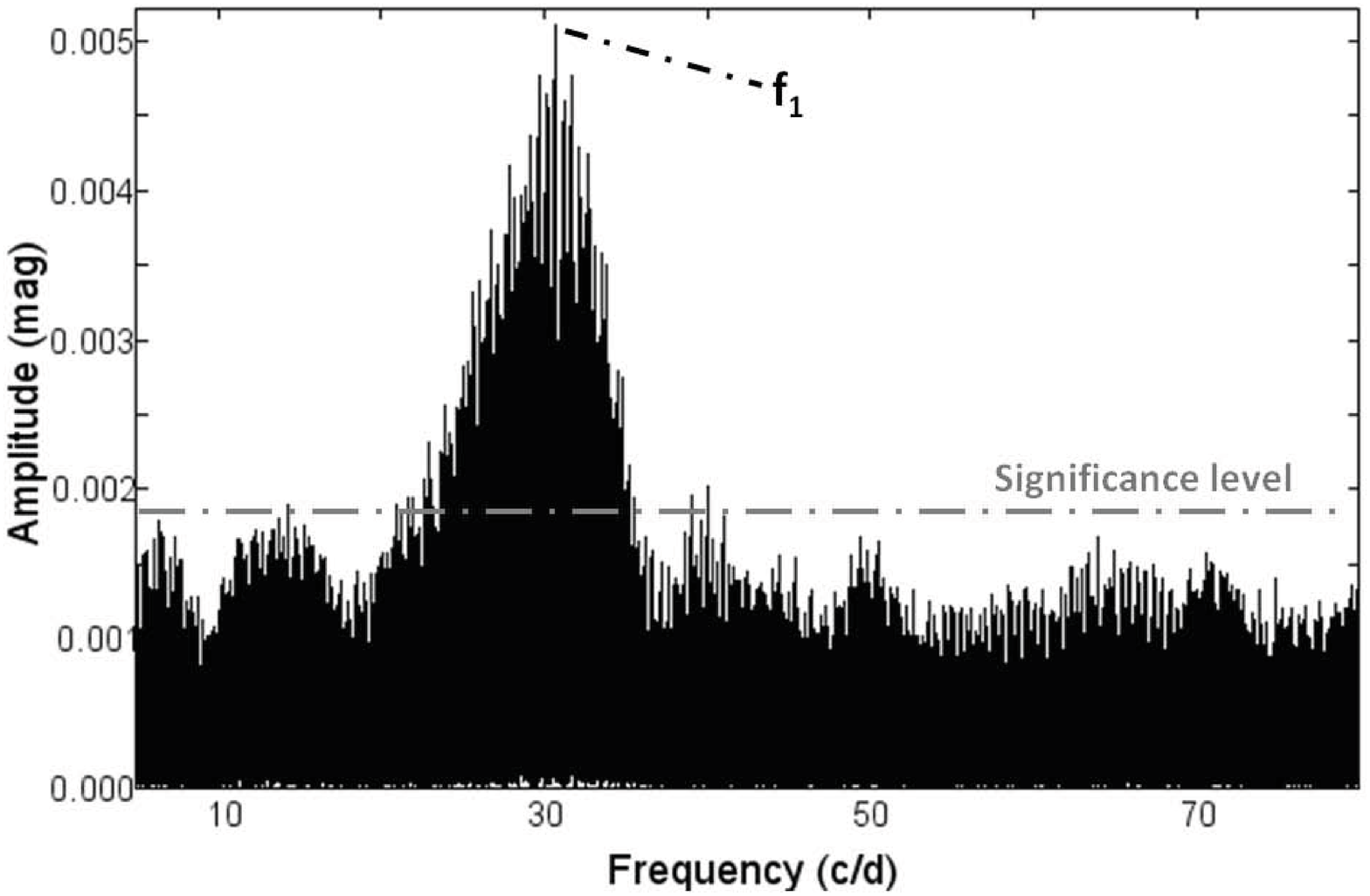}&\includegraphics[width=5.3cm]{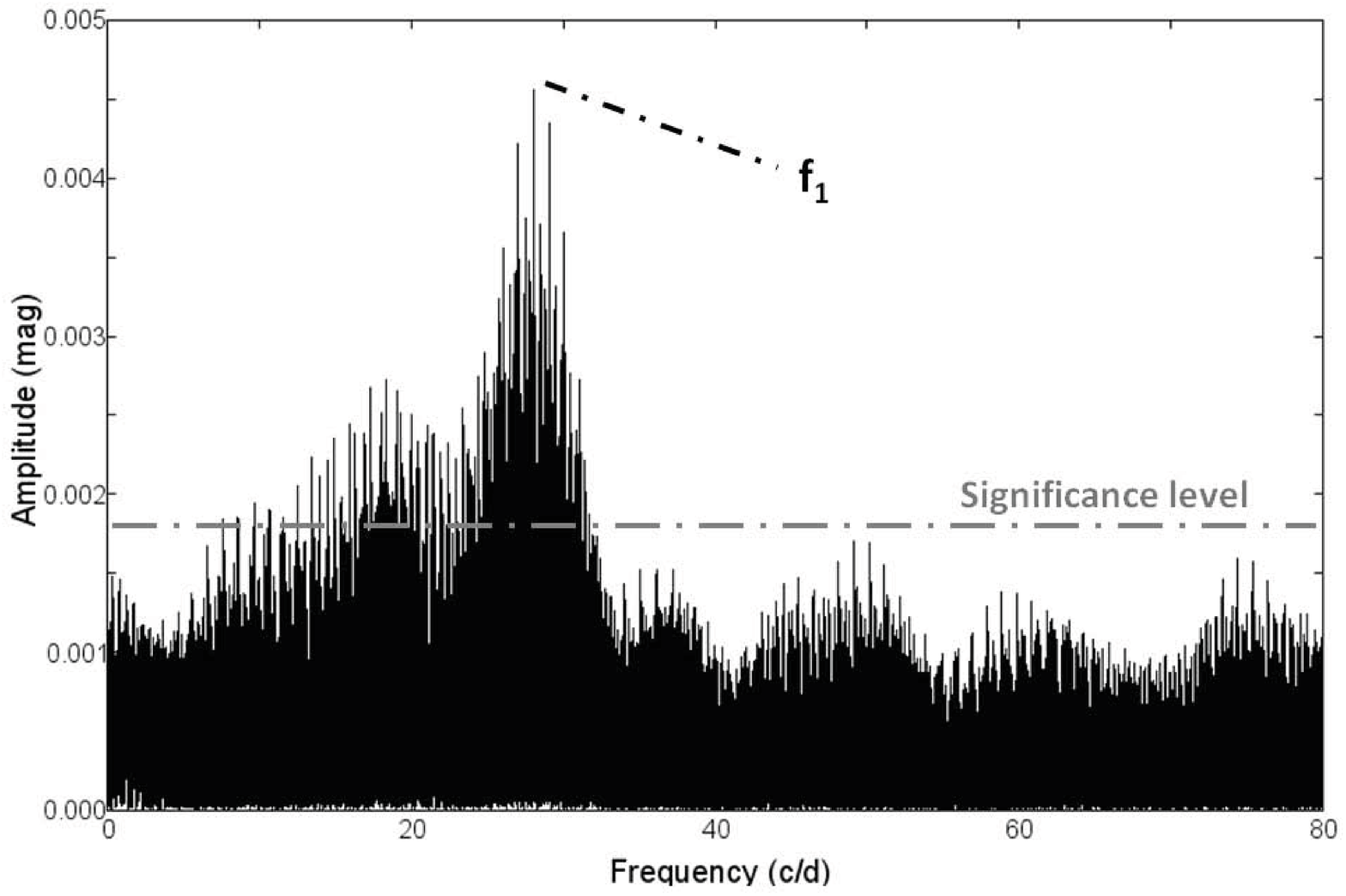}\\
\includegraphics[width=5.3cm]{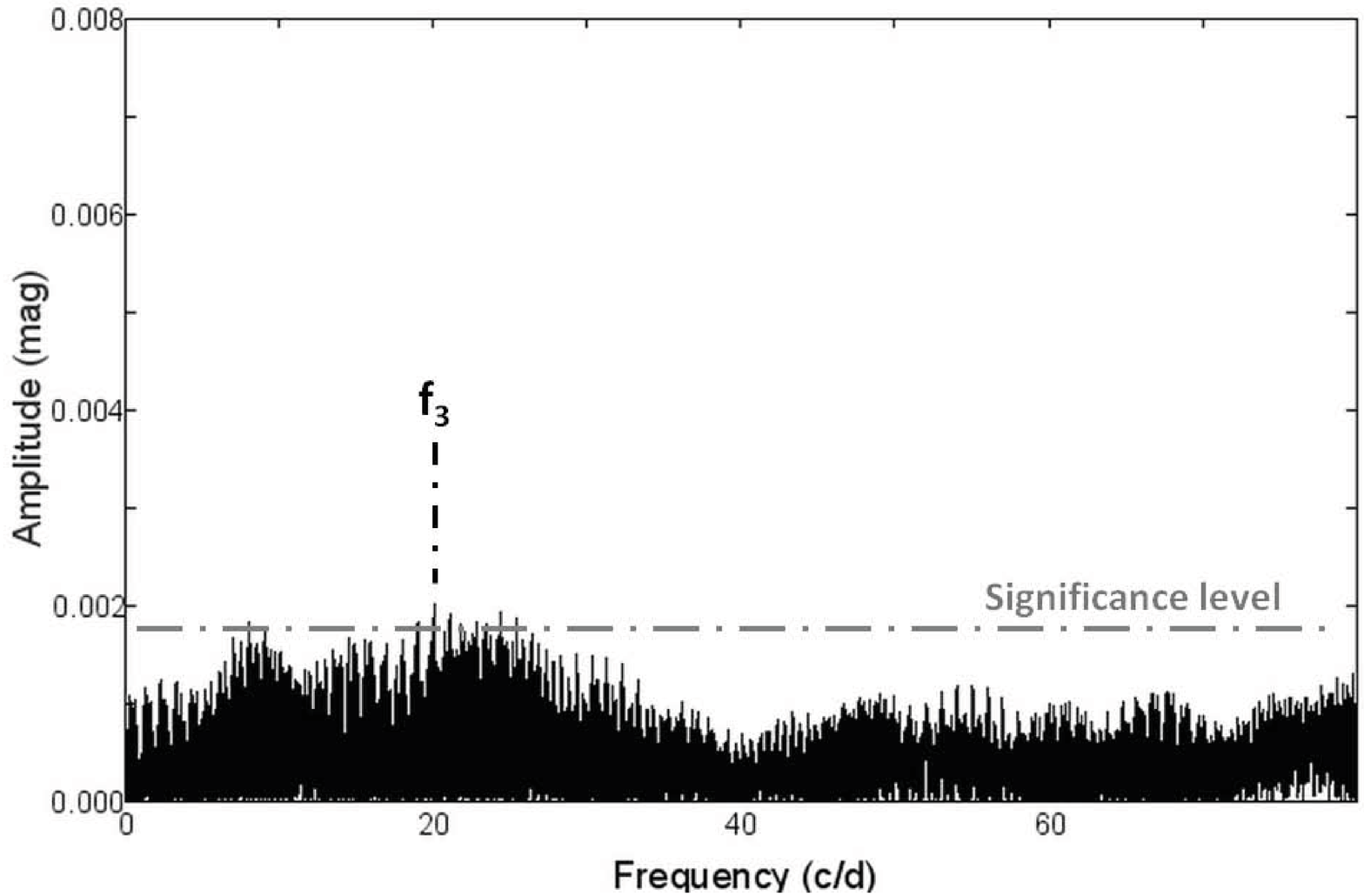}&\includegraphics[width=5.3cm]{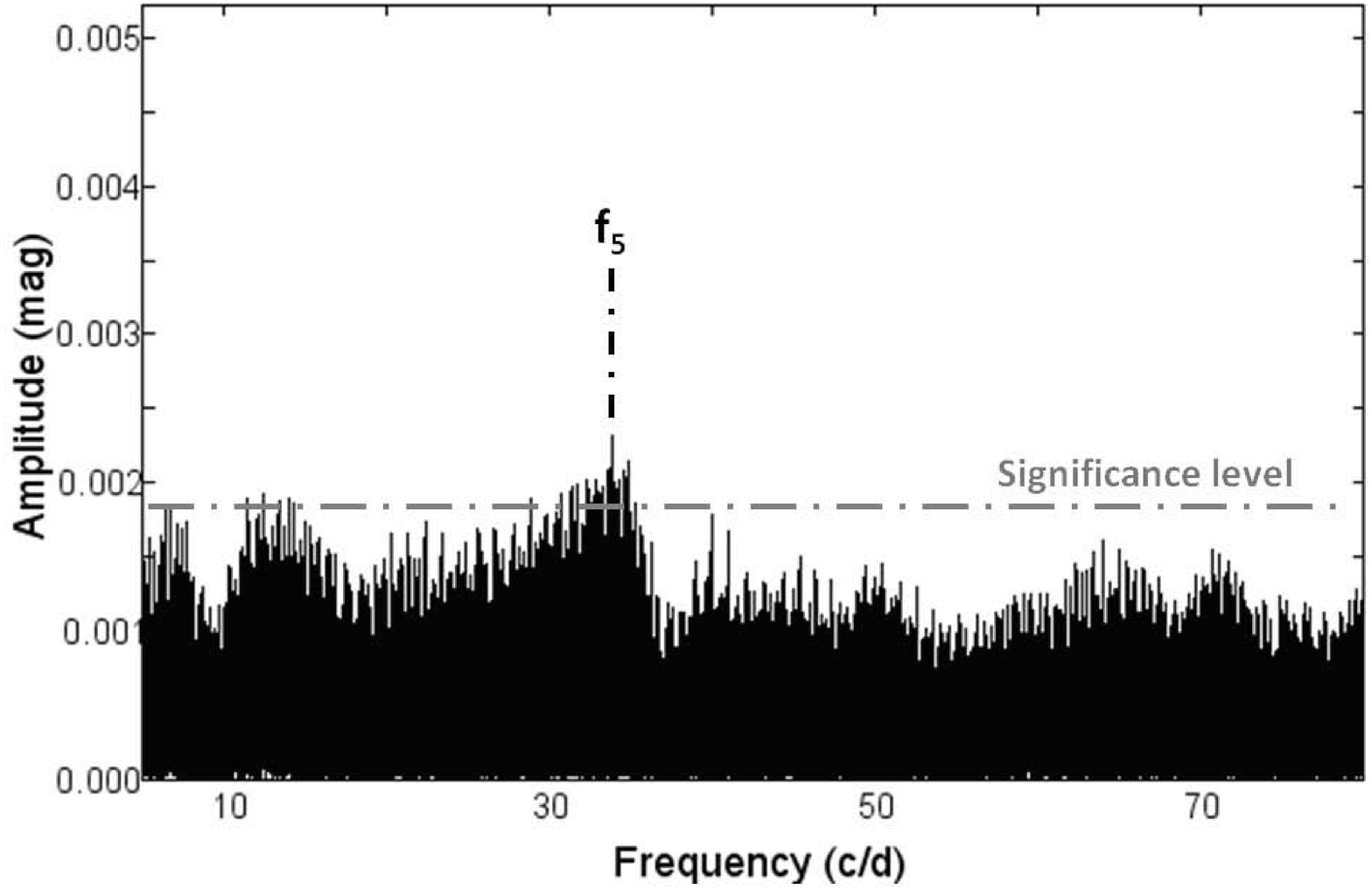}&\includegraphics[width=5.3cm]{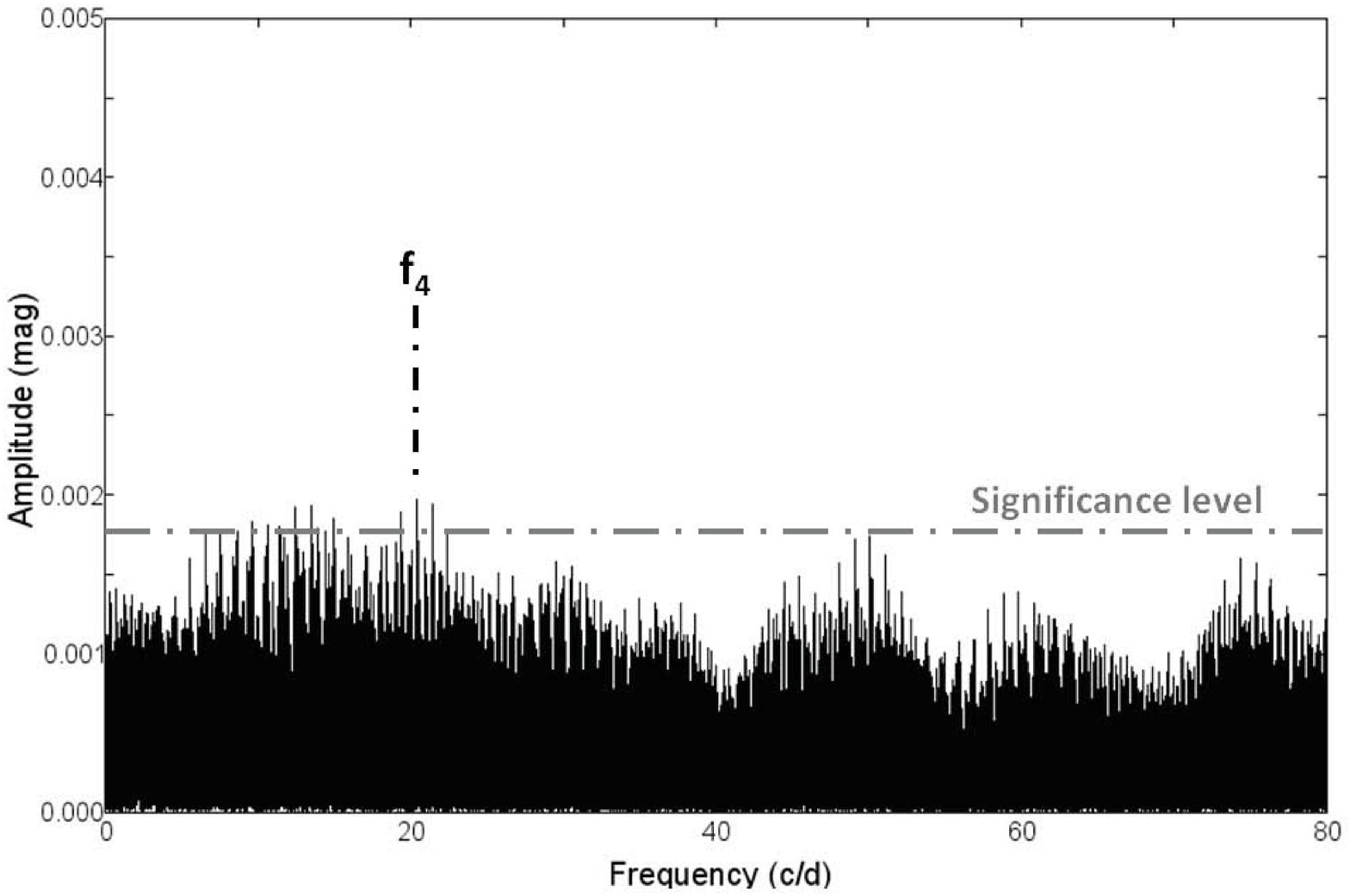}\\
\includegraphics[width=5.3cm]{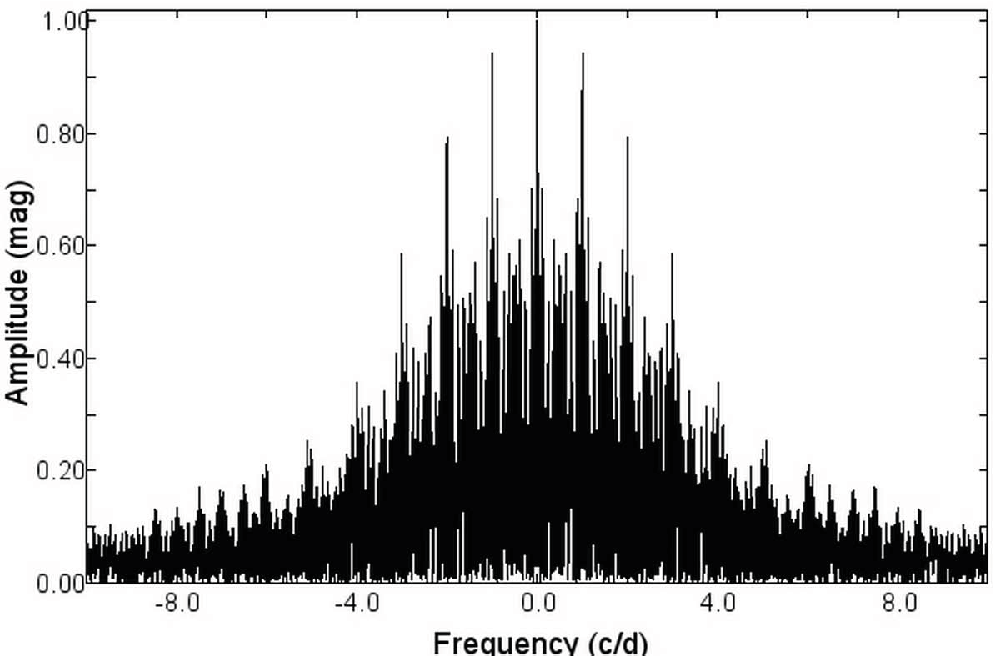}&\includegraphics[width=5.3cm]{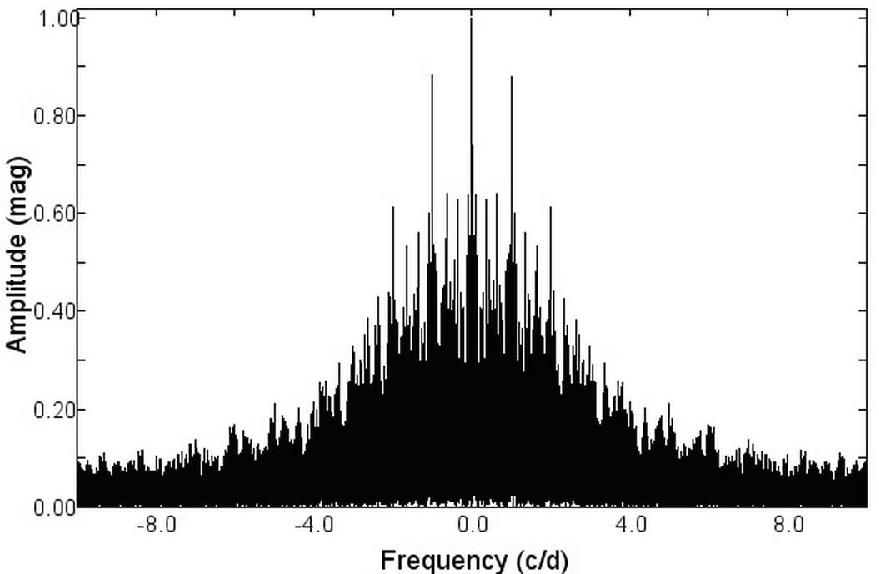}&\includegraphics[width=5.3cm]{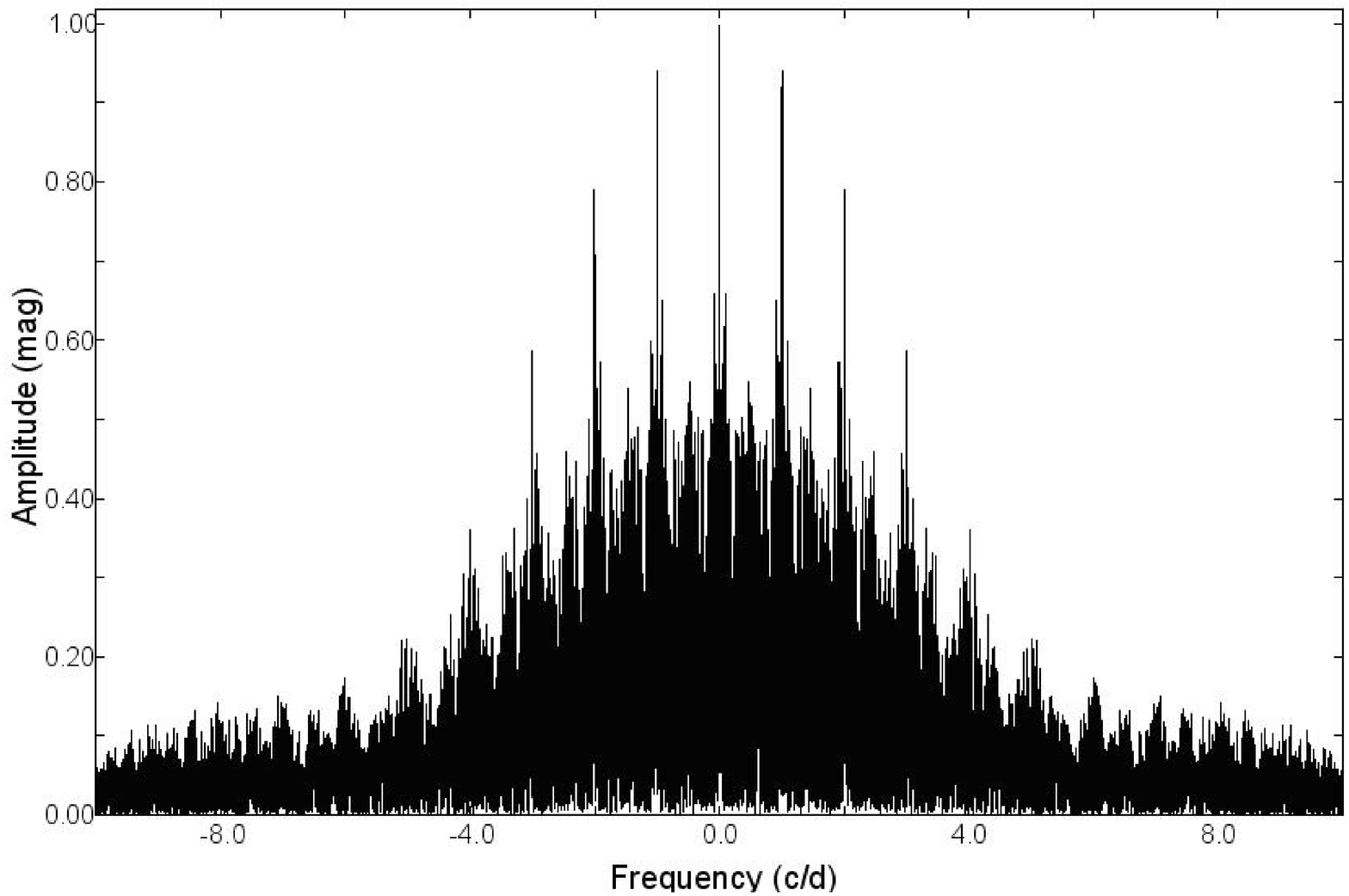}\\
\includegraphics[width=5.3cm]{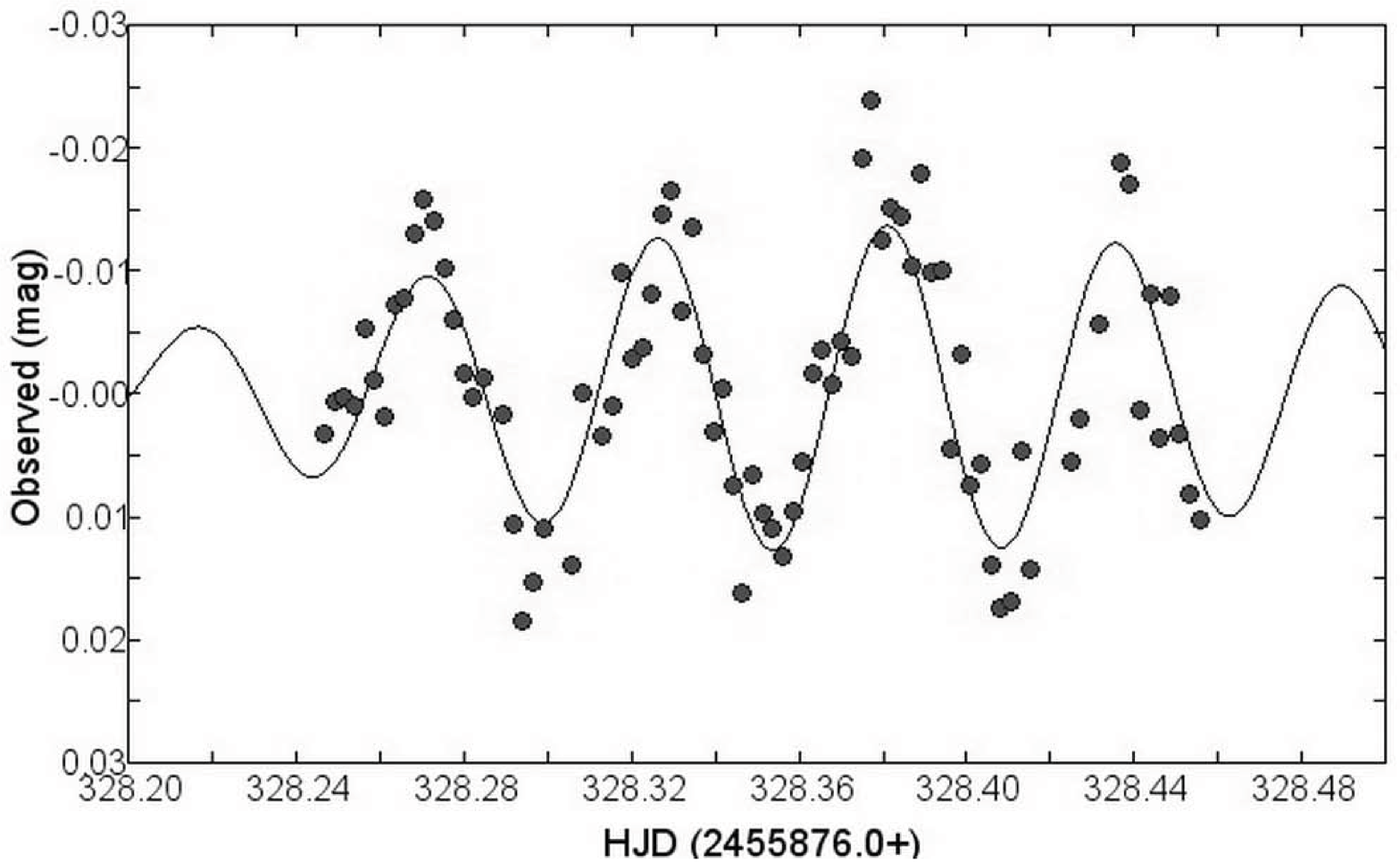}&\includegraphics[width=5.3cm]{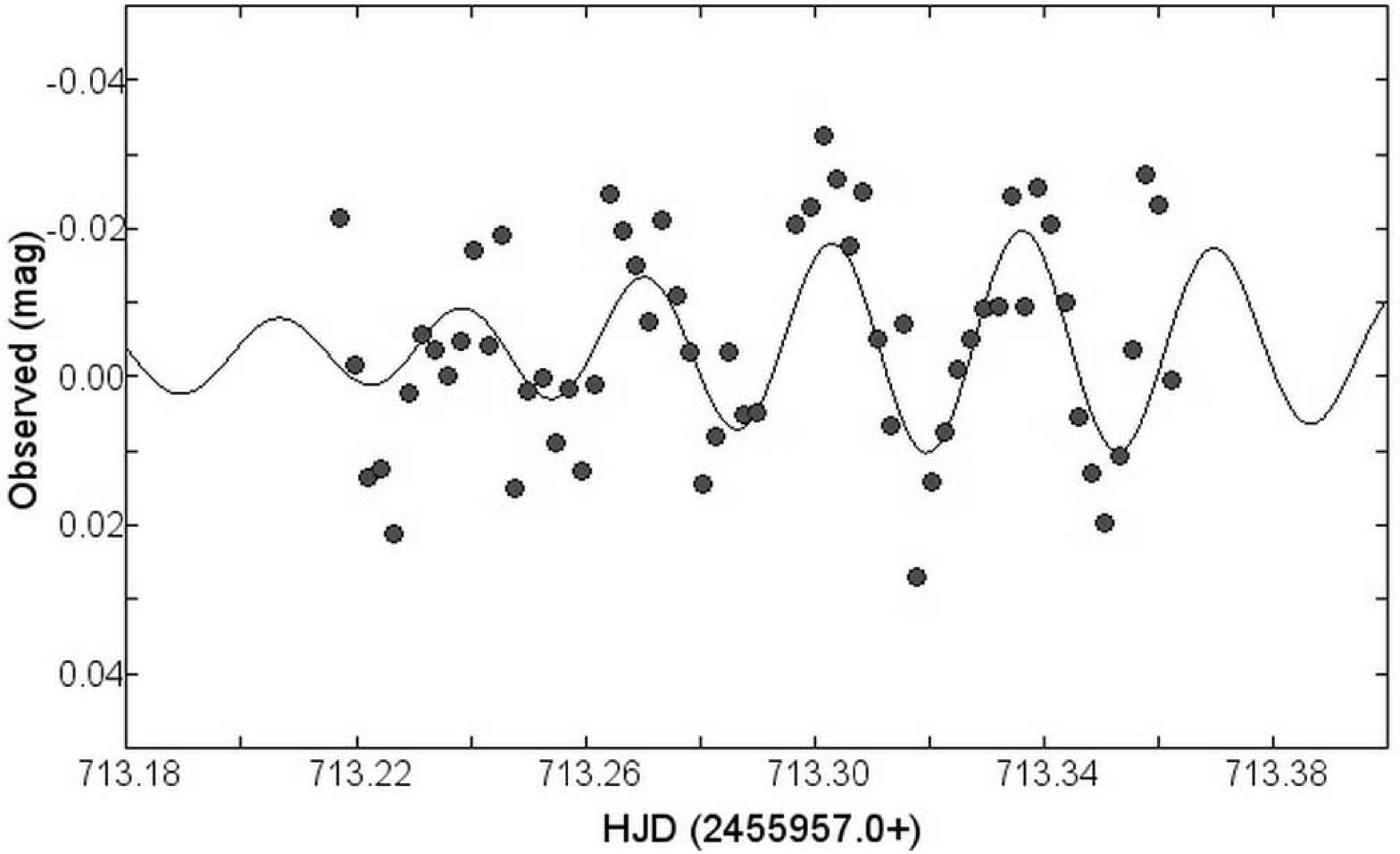}&\includegraphics[width=5.3cm]{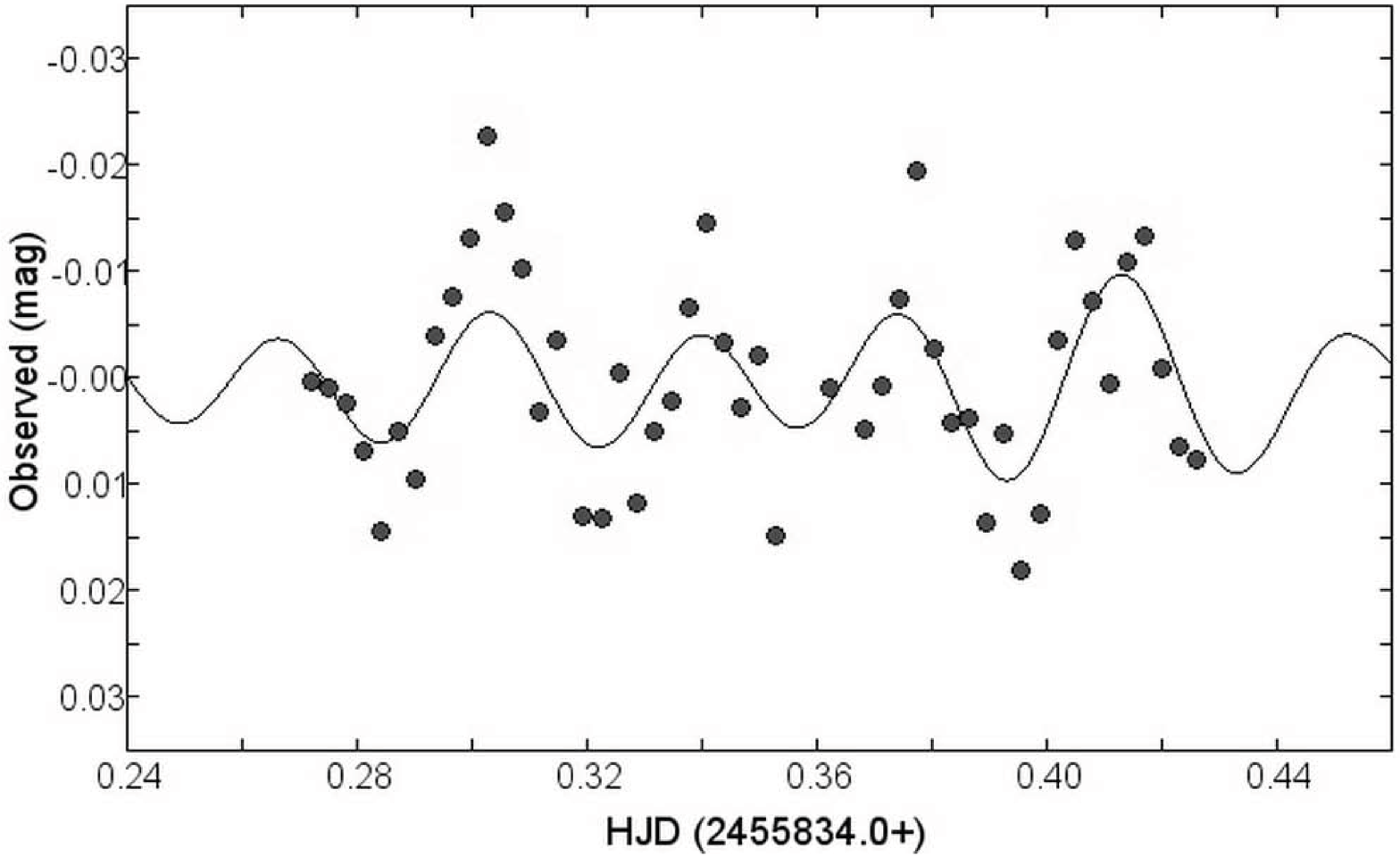}\\
\end{tabular}
\caption{Periodograms of the first (1st row) and the last (2nd row) detected pulsation frequencies inside the $\delta$~Sct frequency range. The significance levels ($\sim0.0018$~mag for USNO-A2.0 0975-17281677 and V729~Aql, and $\sim0.0017$~mag for USNO-A2.0 1200-03937339) are also indicated. 3rd row: Spectral window plots. 4th row: Fourier fits on selected data sets. Each column corresponds to individual studied system.}
\label{figfreqs_dsct}
\end{figure*}

Subsequently, frequency analysis was performed on the residuals of the out of primary eclipse data with the software \textsc{period04} v1.2 \citep{LB05}, that is based on classical Fourier analysis. Given that typical frequencies for $\delta$~Sct stars range between 3-80~c/d \citep{BR00,SO06b}, the analysis was made for this range. After the first frequency computation the residuals were subsequently pre-whitened for the next one, until the detected frequency had a signal-to-noise ratio ($S/N$) less than 4, which is the programme's critical trustable limit. For USNO-A2.0 1200-03937339 and V729~Aql, after the removal of the $\delta$~Sct type frequencies, we performed an additional frequency search in the range 0.3-3~c/d, which is typical for $\gamma$~Dor type pulsations \citep{HF05}, in order to check for possible $\gamma$~Dor-$\delta$~Sct hybrid behaviour. In Fig.~\ref{figfreqs_gdor} are illustrated the periodograms concerning the $\gamma$~Dor frequencies for USNO-A2.0 1200-03937339, in Fig.~\ref{figfreqs_dsct} we plot for all systems the periodograms of the first and last detected frequencies inside the $\delta$~Sct frequency range, the spectral windows, and the fits on selected data sets, while in Table~\ref{tabfreqs} the frequency search results are also given.

\begin{table*}
\centering
\caption{Frequency search results ($A$=semi-amplitude, $\Phi$=phase).}
\label{tabfreqs}
\scalebox{0.83}{
\begin{tabular}{c cccc cccc cccc}
\tableline	
No  &   $f$ (c/d)   &	$A$ (mmag)	&$\Phi~(^\circ)$& $S/N$    &   $f$ (c/d)   &	$A$ (mmag)	&$\Phi~(^\circ)$& $S/N$ &   $f$ (c/d)   &	$A$ (mmag)	&$\Phi~(^\circ)$& $S/N$ \\
\tableline
    &       \multicolumn{4}{c}{USNO-A2.0~0975-17281677}        &        \multicolumn{4}{c}{USNO-A2.0~1200-03937339}     &               \multicolumn{4}{c}{V729~Aql}            \\
\tableline
1   &   18.702 (1)	&		6.5 (3)	&	198 (3)	    &	14.5   &	30.668 (1)	&	   5.1 (4)	&	136 (4)	    &	8.4	&	28.034 (1)	&	   4.2 (4)	&	75 (5)	    &	8.5	\\
2   &   16.625 (1)	&		4.7 (3)	&	121 (4)	    &	8.6	   &	29.528 (1)	&	   4.5 (4)	&	111 (4)	    &	7.7 &	18.292 (1)	&	   2.7 (4)	&	95 (8)	    &	6.6	\\
3   &   20.111 (1)	&		2.3 (3)	&	200 (8)	    &	5.3	   &	26.046 (1)  &	  3.7 (4)	&	112 (5)	    &	6.3	&	25.859 (1)  &	  2.3 (4)	&	14 (9)	    &	5.0	\\
4   &               &               &               &          &	28.930 (1)  &	  3.1 (4)	&	333 (7)	    &	4.9 &	20.406 (1)  &	  2.1 (4)	&	65 (10)	    &	4.9	\\
5   &               &               &               &          &	33.876 (1)  &	  2.4 (4)	&	282 (8)	    &	4.0	&               &               &               &       \\
6   &               &               &               &          &	 0.318 (1)  &	  2.5 (4)	&	283 (9)	    &	7.2	&               &               &               &       \\
7   &               &               &               &          &	 0.933 (1)  &	  1.9 (4)	&	22 (10)	    &	4.6	&               &               &               &       \\

\tableline
\end{tabular}}
\end{table*}

\section{Discussion and conclusions}
\label{DIS}

USNO-A2.0 0975-17281677 is an EB containing a pulsating component with its oscillation frequencies being well inside the typical range of the $\delta$~Scuti stars. According to the $B-V$ index of the system (see section~\ref{INTRO}) and the colour index-temperature correlation of \citet{CO00}, its temperature ranges between 6600~K-7100~K (i.e. early F spectral type star), that is typical for a $\delta$~Scuti star. Its photometric model was not presented due to the poor coverage of its LC, however, its LC is of Algol-type, but its Roche geometrical configuration could not be certified. The hotter component (primary) is the one which is eclipsed during the primary minimum and probably this is the pulsating star of the system. It oscillates in a triple-frequency mode with the dominant one to be $\sim18.7$~c/d. The system can be categorized either as an oEA star (if it is semi-detached) or as a detached system with a $\delta$~Scuti member.

Both the spectral type and the pulsation frequencies of USNO-A2.0 1200-03937339 lead to the conclusion that the primary star is a $\delta$~Scuti star, which pulsates in five frequencies with the dominant one $\sim30.7$~c/d. Two more frequencies (i.e. $f_6$ and $f_7$) in the pulsation frequency range of $\gamma$~Dor type stars were found and they are discussed in the following paragraphs. Moreover, according to its conventional semi-detached Roche configuration can be clearly categorized as an oEA star. The secondary component is more evolved than the primary, it has already lost fraction of its initial mass, and up to date continues to transfer mass to the primary.

The primary component of V729~Aql, according to its temperature and the pulsation frequencies values, is a $\delta$~Scuti star pulsating in four oscillating modes. The dominant pulsation frequency has a value 28.034~c/d, while it must be noticed that the ratio $f_{4}/f_{1}$ is $\sim0.73$, which is very close to the value 0.75, that is typical for the radial fundamental and first overtone modes. Similarly to USNO-A2.0 1200-03937339, this system is definitely an oEA star, since its secondary component was found to fill its Roche lobe, it has already transferred mass to the primary and probably still continues to.

\begin{figure}[h]
\includegraphics[width=7.5cm]{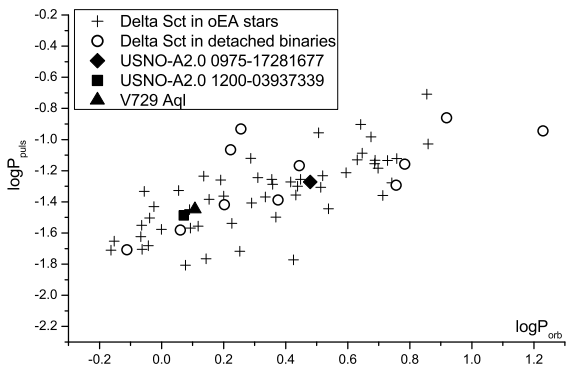}\\
\caption{The positions of the primary components of the studied systems in the $P_{\rm orb}$-$P_{\rm puls}$ diagram along with the $\delta$~Scuti stars-members of oEA stars and of detached binaries.}
\label{figPP}
\end{figure}

The pulsating components of the studied systems are presented in the $P_{\rm orb}$-$P_{\rm puls}$ diagram with other stars having similar properties for direct comparison (see Fig.~\ref{figPP}). The data for this diagram were taken from \citet{LI12}. In this plot it is shown that the pulsators of the studied systems lie well inside the periods' range of the EBs with a $\delta$~Scuti star-member. Based on the work of \citet{ZH13}, we calculated the dominant pulsations constants $Q$ and the $P_{\rm puls}/P_{\rm orb}$ ratios for USNO-A2.0 1200-03937339 (0.0123 days and 0.0276, respectively) and V729~Aql (0.0159 days and 0.0278, respectively). According to these values, these systems follow well the distribution $P_{\rm puls}/P_{\rm orb}$ vs. $q$ of EBs with $\delta$~Scuti component as drawn by \citet{ZH13}, while the $Q$ values of their primaries (which are $<0.033$~days) show that they should pulsate in p-modes. Moreover, we calculated the $P_{\rm puls}/P_{\rm orb}$ ratio for USNO-A2.0 1200-03937339 but for the pulsation frequency $f_6$, in order to compare it with other $\gamma$~Dor-$\delta$~Sct hybrid stars-members of EBs. However, according to \citet{ZH13} the sample of such hybrids presently consists of at most two hybrid pulsator eclipsing binaries, namely CoRot100866999 and possibly KIC4544587. Their $\gamma$~Dor $P_{\rm puls}/P_{\rm orb}$ ratio of 0.223 and 0.227, respectively, differ by an order of magnitude from the value of 2.666 that we found for USNO-A2.0 1200-03937339. Therefore, perhaps $f_6$ and $f_7$ result from observational drifts or imperfect LC fitting rather than being intrinsic to the star. Certainly, this issue remains open until the statistical sample of this type of hybrid pulsator becomes significantly larger.

We strongly suggest the multi-filtered photometric monitoring of these systems for: a) verification of the present results, especially for the low amplitude frequencies, b) the detection of more pulsation frequencies, and c) the derivation of the excitation modes (i.e. $l$-degrees). Moreover, for USNO-A2.0 1200-03937339 short-time continuous observations (i.e. order of weeks) are proposed in order to certify the existence of the lower frequencies, and for USNO-A2.0 0975-17281677 systematic observations aiming for complete coverage of its LCs are suggested in order to find its geometrical characteristics. Finally, spectroscopic observations are also suggested not only for the pulsations, but also for the components' radial velocities curves, which will lead us to calculate their absolute properties (i.e. masses, radii) with high accuracy.

\section*{Acknowledgments}
This work was performed in the framework of PROTEAS project within GSRT's KRIPIS action for A.L., funded by Hellas and the European Regional Development Fund of the European Union under the O.P. Competitiveness and Entrepreneurship, NSRF 2007-2013. In the present work the SIMBAD database, operated at CDS, Strasbourg, France, and Astrophysics Data System Bibliographic Services (NASA) have been used. This work has also made use of SDSS (http://www.sdss.org/). Funding for the SDSS and SDSS-II has been provided by the Alfred P. Sloan Foundation, the Participating Institutions, the National Science Foundation, the U.S. Department of Energy, the National Aeronautics and Space Administration, the Japanese Monbukagakusho, the Max Planck Society, and the Higher Education Funding Council for England. We thank the anonymous referee for his/her valuable comments, which improved the quality of the present paper.

\end{document}